\documentclass[mathleft
]{an}
\usepackage{graphicx}
\usepackage{times}
\begin{document}

% The following seven commands are intended for editorial usage and should be ignored by
% the author(s).
\Pagespan{789}{}% Document's page range. 
% If second parameter is left empty, the last page is computed automatically.
\Yearpublication{2006}%
\Yearsubmission{2005}%
\Month{11}%   
\Volume{999}%  
\Issue{88}% 
% \DOI{This.is/not.aDOI}% 

\def\aap{{\it Astr.~Ap.}}     %Astronomy & Astrophysics%
\def\nat{{\it Nature}}      %Nature%
\def\et{{et\thinspace al}}    %et al.%

\title{Transient Detection and Classification}

\author{A.C.Becker\inst{1}\fnmsep\thanks{Corresponding author:
  \email{becker@astro.washington.edu}\newline}
}
\titlerunning{Transient Detection and Classification}
\authorrunning{A.C. Becker}
\institute{
University of Washington, Department of Astronomy, Box 351580, Seattle WA 98195-1580
}

\received{}
\accepted{}
\publonline{later}


\abstract{ I provide an incomplete inventory of the astronomical
  variability that will be found by next--generation time--domain
  astronomical surveys.  These phenomena span the distance range from
  near--Earth satellites to the farthest Gamma Ray Bursts.  The
  surveys that detect these transients will issue alerts to the
  greater astronomical community; this decision process must be
  extremely robust to avoid a slew of ``false'' alerts, and to
  maintain the community's trust in the surveys.  I review the
  functionality required of both the surveys and the telescope
  networks that will be following them up, and the role of VOEvents in
  this process.  Finally, I offer some ideas about object and event
  classification, which will be explored more thoroughly by other
  articles in these proceedings.  }

\maketitle

\section{Introduction}
Next generation surveys such as
Pan-STARRS\footnote{http://pan-starrs.ifa.hawaii.edu} and
LSST\footnote{http://www.lsst.org} promise to open up the time domain
of astronomical variability to the general community as a service.
This will allow the global study of all on--going phenomena in
real--time, enabling both the small aperture amateur astronomer and
the large aperture professional.  It also places great responsibility
on the surveys themselves to provide a reliable stream of information.

The scope of the planned data release is unprecedented; 10$^{5-6}$
transient ``alerts'' are predicted to be generated on a nightly basis
by the LSST alone.  This will place a huge burden on follow--up
networks in the very near future.  Undertaking follow--up of these
alerts could easily consume {\it all} of the available global
telescope resources unless intelligent decisions are made about which
events to focus on.  The volume of the data streams will preclude this
decision from being made by a human.

To enable intelligent, autonomous follow--up systems, the VOEvent
protocol has been developed as a means of automatically conveying
information between astronomical resources.  To take advantage of this
stream, Heterogeneous Telescope Networks (HTNs) are being implemented
to undertake follow--up of alerts.  This purpose of this conference is
to address the practical implementation of this marriage.  This paper
will address the types of astronomical variability the surveys will
have to sort through, the types of alerts that the surveys will
generate, and present various ideas on the classification of these
alerts.  I will emphasize that the surveys must do more than generate
alerts; they must also listen to the follow--up results of the
community or risk retaining outdated classifications based upon their
own limited data.

\section{Sources of Astronomical Variability}
The exciting part about surveys such as LSST is that they intend to
detect, classify, and release information on {\it all} variability
found whether it be photometric or astrometric.  This provides a
technical challenge for the surveys in terms of autonomy and
reliability that has been approached but not yet demonstrably met in
precursor efforts (Bailey \et. 2007; Becker \et. 2004).

Modern surveys use image subtraction techniques to remove the static
portion of their images, leaving only the residual flux of objects
that have varied in brightness or position.  In these difference
images, the astronomical signal is almost exclusively objects
elongated due to astrometric motion, or positive or negative point
sources that have varied in brightness or position.  I provide below
an incomplete listing of astronomical phenomena expected to be found
in these images, starting with the foreground of astrometrically
variable objects and ending with the most distant of cosmological
explosions.

\subsection{Astrometric Variability}
Earth--orbiting satellites provide the least interesting (for most
scientists) but most destructive foreground.  They move at an angular
velocity of order $10^4 \arcsec$ s$^{-1}$, and completely traverse a
single field--of--view in a typical exposure time.  Their image
signature is a nearly infinitely elliptical streak.  Inactive
satellites may also be tumbling, which yields a periodic sequence of
flashes along the satellite's vector.  Most photometry software is not
designed to accurately model such a trail, and thus will deblend the
trail into numerous elongated sources.  This sequence of detections
may be identified in a database using e.g. a Hough transform (Storkey
\et. 2004).

Solar system objects move with an apparent angular velocity that is a
combination of their own spatial velocity (the dominant term for
nearby objects) and the Earth's reflex motion (dominant for distant
objects).  These objects may appear elongated depending on their
apparent angular motion, angle from opposition, and exposure time.
%Elongated objects have the profile of the image's point spread
%function (PSF) in the dimension perpendicular to motion, and the PSF
%convolved with a line along the direction of motion.  
Near--Earth objects (NEOs) have closest approaches of order 0.001 AU,
where their apparent sky motion is as fast as $1 \arcsec$ s$^{-1}$.
Main belt asteroids are found between 1 and 2 AU from Earth, orbiting
between Mars and Jupiter.  Typical angular velocities are $10 \arcsec$
hr$^{-1}$.  Trans--Neptunian objects (TNOs), or Kuiper belt objects
(KBOs), orbit within or beyond the orbit of Neptune near 40 AU.
Moving at $\sim 1 \arcsec$ hr$^{-1}$, they appear as single--epoch
point sources in all but the longest astronomical exposures, and are
difficult to distinguish from background transients.  An ensemble of
moving objects imaged nightly will yield a ``new'' detection per
object per image.  These single--epoch detections may be efficiently
linked in\-to orbits (Kubica \et. 2007), with any orphaned detections
potentially background variability.  Main Belt asteroids and TNOs are
concentrated strongly in the ecliptic plane, which may be used as a
prior to disentangle them from background cosmological transients.
Beyond the Kuiper belt, there are very few Solar System objects known.
The Oort cloud of comets is thought to exist between $10^4$ and $10^5$
AU, however no objects are currently known at this distance.  This
transition region reflects the boundary from objects that primarily
reflect the Sun's radiation to objects that produce their own.

Beyond this regime, astrometric motion is only noticeable in the
nearest or most rapidly moving of our stellar neighbors.  The extreme
example of this is Barnard's Star, whose apparent angular motion is
$\sim 10 \arcsec$ yr$^{-1}$.  The difference imaging signature of high
proper motion objects is a dipole whose nodes grow more separate as a
function of time, over the timescale of years.  Since dipoles are also
a classical signature of image subtraction failure, these may be
difficult to distinguish from systematic noise until an ensemble of
difference images is examined.

\subsection{Photometric Variability}
Beyond the solar neighborhood, all variability will have the higher
order moments of the image's PSF (a noteworthy exception are supernova
light echoes, which leave large, low surface brightness features
evolving radially from a central point; Rest \et. 2005).  Since these
variables all appear as point sources, contextual and temporal
information are required to classify the nature of each event.  I
summarize some of the expected source populations below.

Planetary transit searches are undertaken on nearby, apparently
bright, stars out to a distances of a couple of kilo--parsecs
(Pollacco \et. 2006).  Their lightcurves are characterized by minute
(several percent), periodic (several days) decrements in the host star
brightness as it is transited by its exoplanet.

At similar distances, a foreground ``fog'' of flaring M--dwarf stars
has been found (Kulkarni \& Rau 2006).  This comprises low--mass stars
that are too faint to be seen in a single image but which may flare
suddenly, yielding short timescale ($\sim 1000$ s) apparently hostless
transients (Becker \et. 2004).  The all--sky rate of these may be up
to tens of millions per year, making them a significant foreground to
cosmological variability.

At a distance of 8 kpc, the Galactic center is used to backlight a
foreground microlensing population, yielding (typically) symmetric,
unique source brightenings (Paczynski 1991).  Durations of these
events can be anywhere from $10^0$ to $10^3$ days.  Because of their
angular coverage and temporal sampling, microlensing surveys have also
yielded a wealth of information on stellar variability of all types.
At around 50 kpc, the Large and Small Magellanic Clouds are targeted
for microlensing by objects in our Galactic halo (Paczynski 1986),
with typical timescales of $10^2$ days.  Individual stars in these
galaxies are able to be resolved from the ground and photometered,
although the blending can be quite severe.

RR Lyrae are periodically pulsating horizontal branch stars that may
be recognized from their lightcurve shapes.  Their periods can range
from 0.2 to 2 days.  Because of their well--studied period--luminosity
relationship, their apparent magnitude distribution can be used to
infer Galactic structure (Sesar \et. 2007).  They have been found in
our Galactic halo out to 100 kpc; LSST expects to probe Galactic
accretion structure using RR Lyrae out to 400 kpc.

At a distance of 1000 kpc, M31 is also targeted by microlensing
surveys, but the lensed stars are unable to be resolved and difference
imaging techniques are necessary (Uglesich \et. 2004).

The HST Key Project measured lightcurves of Cepheid variable
stars in nearby galaxies out to 10 Mpc to measure the Hubble constant
(Freedman \et. 2001).  Cepheids are intrinsically several magnitudes
brighter than RR Lyrae, meaning they can be seen to larger distances.
They also have well calibrated period--luminosity relationships, with
periods of 5 to 30 days.

Starting at around 100 Mpc (a redshift of z $\sim$ 0.1), a wealth of
nearby supernovae have been observed, starting with the Calan-Tololo
sample of Type Ia supernova (Hamuy \et. 1996).  Because supernovae are
typically enmeshed in the light of their host galaxies, image
subtraction techniques are required to photometer the supernova light
uniquely.  Contextually, supernovae are commonly (but not always)
found near an extended host galaxy.  Their lightcurves experience a
steep rise ($\sim 20$ days) and a gradual fall over $\sim 100$ days
characteristic of their subtype (Ia, Ib/c, IIp, etc...).

Just beyond this sample, around a redshift of 0.1, the closest Gamma
Ray Bursts have been found (GRB 031203; Gotz \et. 2003).  Their
temporal evolution is much faster than supernovae, rising and falling
within hours to days, making optical discovery of these phenomena
extremely di\-fficult.

Medium redshift supernova surveys like the SDSS--II Supernova Survey
find events out to z $\sim 0.4$ (Frieman \et. 2007).  Higher redshift
supernova surveys, including both ESSENCE (Wood-Vasey \et. 2007) and
the Supernova Le\-gacy Survey (SNLS; Astier \et. 2006), detect Ia events
between 0.3 $<$ z $<$ 1.0, while the highest redshift supernova
surveys involving the HST find Ia supernovae out to $z \sim 1.4$
(Riess \et. 2007).  At their most distant, supernova are only visible
for several days around their peak.
Finally, the highest redshift cosmological transient, GRB 050904, was
found at $z \sim 6.3$ (Haislip \et. 2005).

The overall extent of astronomical variability is clearly enormous,
and individual surveys have typically been commissioned to address a
subset of the above whole.  However, surveys such as LSST anticipate
not only {\it detecting} and {\it classifying} the above phenomena,
but doing so in real--time.

\section{Alerts and Classification of Sources}
The general paradigm for alert generation is that a new event will be
recognized by a real--time survey pipeline, and the survey will
subsequently release an ``alert'' describing the detection (and
possibly supporting characterization) observations.  The format of
these alerts will be as VOEvent packets.  The decision to take action
based upon a VOEvent is undertaken by ``intelligent agents'' as a part
of each HTN.  Depending upon the science goals of each HTN, different
agents will make different decisions based upon the same information.
These actions will depend upon the event classifications of the survey
and potentially of the agent itself.  I summarize below some of the
requirements necessary for this model to succeed.

\subsection{Alerts and Followup}
A VOEvent packet includes {\tt inference} fields, where the survey
lists to the best of its abilities the classification of the event, as
well as the {\tt probability} that the event is of this class.  These
alerts may be {\it urgent} in nature, suggesting immediate follow--up.
The HTN resources decide to target or not based on the {\tt inference}
and {\tt probability}, and their particular science goals.  The
surveys and follow--up networks will also be releasing more prosaic
{\it informational} alerts that do not necessarily require action.
These should be released each time an object that has been alerted on
has been followed up.

Each HTN's intelligent agent may be assumed to be autonomous from the
others.  Consequently, these different agents may come to different
conclusions about the true nature of the event given the same
information.  This could easily lead to asynchronous/conflicting
evolution of knowledge between networks.  

The instantaneous state of knowledge about an alert can be extracted
from its ensemble of VOEvents by a citation mechanism that links
multiple observations together -- they are federated by mutual
citation.  As an alternative, the concept of a ``broker'' has been
introduced representing an agent who centralizes and disseminates this
information.  As with the surveys, the broker's survival requires
engendering trust from its subscribers.  Surveys may play a hybrid
role in this model, releasing VOEvents on the entirety of their alerts
but also serving as brokers by releasing more descriptive alerts
(including e.g. fit parameters) on particular subsets of events.

\subsection{Source Classification}
To release accurate alerts with a minimum of false positives, i.e. to
have trustworthy {\tt inference} in VOEvents, model templates of event
behavior must be built beforehand.  To first order, there are two
levels of classification requiring image--based (spatial) and
lightcurve--based (temporal) models.

\subsubsection{Spatial Classification}
The morphological classification of flux in an image is a
well--studied problem.  All astronomical images have a characteristic
point spread function (PSF), which is the transfer function of a point
source through the atmosphere, telescope optics, and detecting
instrument.  Objects sharper than the PSF are not likely to be real
astronomical phenomena; objects broader than the PSF may be noise
artifacts or resolved objects such as galaxies or comets.  Moving
objects will have the profile of the PSF in the dimension
perpendicular to motion, and the PSF convolved with a line along the
direction of motion

Classification in difference images occurs through comparison of the
residual flux with the PSF profile.  A spatial model of the PSF is
typically built from the data itself.  This step requires direct
access to the images where the variability was detected.  Given the
potential bandwidth and disk access requirements of the alternative,
this step is almost exclusively the responsibility of the survey.

\subsubsection{Temporal Classification}
Sets of detections may be linked into a lightcurve through mutual
citation or by a broker.  To recognize a given event as a certain
class of phenomena, data models must first be built that span the
parameter space characterizing the phenomena.  The lightcurve data are
then compared to each template lightcurve in a probabilistic sense,
determining which model (if any) they fit best.  These ideas are
explored more fully in other articles in these proceedings (Bloom,
Mahabal, Bailey).

A complication for these next generation surveys is that a data model
incomplete at the e.g. $1\%$ (or $0.1\%$, or even $0.01\%$) level will
result in an unacceptably large number of falsely classified alerts.
The models need to be nearly bullet--proof.  To build these models,
both theoretical and experimental priors should be taken into account.
However, the observational data best span the range of actual (as
opposed to expected) phenomenology.  This suggests that an ideal use
of existing datasets is to build these models {\it before} they are
required by LSST and Pan-STARRS.  A prime example of phenomenological
model building is in the description of supernova Ia lightcurves
(e.g. Guy \et. 2007).

One successful implementation of lightcurve classification driving
real--time decision making is by the SDSS-II Supernova Search (Sako
\et. 2007).  The evolving light\-curves are fit to various supernova
lightcurve models after each epoch of observation.
%Currently, spectroscopic follow--up is required to confirm the
%redshift and type of each supernova; since the survey is searching for
%Type Ia events, the sample targeted for spectroscopic follow--up
%should have a high a fraction of Ia events as possible.  
SDSS-II find that after 2--3 lightcurve epochs, approximately $90\%$
of objects {\it photometrically} classified as Ia end up being Ia.
While their analysis does not examine the efficiency of this process
(how many Ia are missed), it is nevertheless an encouraging precursor
effort.

\subsection{Data Access and Computational Requirements}
In the process of alerting and event classification, there are
multiple stages of computation and inference undertaken by members of
this community.  The separation of responsibilities emerges by
examining which aspects of the data are required to address each
problem, and who has easiest access to it.

As concrete examples, plausible alerts on moving objects include : a
known object may soon be lost, or its error ellipse is growing at an
unacceptable rate; an object is brighter/fainter than past behavior
would indicate; or the likelihood of this object impacting Earth is
significant.  Each of these alerts requires computation and knowledge
exchange.  For the first, the future cadence of the survey is
required; this sort of computation is best done by the survey.  For
the second, the past behavior of the object must be accessible or
queryable.  This is a requirement on the survey or on the federation
of all data on the object.  The last use case requires significant
computational resources to project all moving objects into the future.
This computation is most likely to get done (with high latency) by a
motivated user.

In the use case of a newly detected transient, it is unclear who (if
anyone) will have the final say in classification of the event.
Especially in the early portion of the lightcurve, the inference will
be changing rapidly as new data come in.  A broker seems most useful
at this critical stage, tying all data together into a coherent
inference.

Finally, there are the cases of known objects whose current behavior
is unexpected.  Examples include exotic deviations in an on--going
microlensing event due to a planet orbiting the lens, stellar flares
from low--mass stars, and deviations in the timings of transiting
systems due to the gravitational influence of other unseen planets.
In these cases, real--time lightcurve fitting is required (an
extensive task), not likely to happen by the survey unless it also
chooses to serve as a broker.

\section{Conclusions}
I have detailed some of the interplay between surveys and follow--up
networks in time--domain astronomy, highlighting their potential roles
and responsibilities.  While the concept of data federation using
VOEvent's citation mechanism may work, it may also lead to different
inferences by different resources.  An alternative is to broker the
evolving ``truth'' regarding an event by trusted agents in the system.
These agents have the potential to fully realize (or scuttle) the
successful interplay between surveys and follow--up HTNs.

The surveys and agents each play a role in the ultimate classification
of each event.  I emphasize that the surveys should also {\it listen}
to the network, and decide if and how to allow these external
resources to influence their internal event classifications.  Finally,
the need for accurate and precise data models is commensurate with the
data flow that will be compared against them.  These models should be
built sooner rather than later using currently existing data sets, to
ensure that the promise of the first several years of these surveys is
not lost to faulty or inaccurate alerts.

%\begin{thebibliography}{}
%  \bibitem{} Author1, A.B., Author2, C.D.: 2001, AN 322, 1
%  \bibitem{} Author3, E.F., Author4, G.H.: 2001, AN 322, 10
%  \bibitem{} Author5, I.: 2001, AN 322, 20
%  \bibitem{} Author6, J.: 2001, AN 322, 30
%\end{thebibliography}

%\bibliography{ms}

\begin{thebibliography}{10}

\bibitem{2006A&A...447...31A}
P.~{Astier} \et. : 2006,
\newblock {\em \aap} 447, 31--48

\bibitem{2007ApJ...665.1246B}
S.~{Bailey} \et. : 2007,
\newblock {\em \apj} 665, 1246--1253

\bibitem{2004ApJ...611..418B}
A.~C. {Becker} \et. : 2004,
\newblock {\em \apj} 611, 418--433

\bibitem{2001ApJ...553...47F}
W.~L. {Freedman} \et. : 2001,
\newblock {\em \apj} 553, 47--72

\bibitem{2007arXiv0708.2749F}
J.~A. {Frieman} \et. : 2007,
\newblock {\em ArXiv e-prints} 0708.2749

\bibitem{2003GCN..2459....1G}
D.~{Gotz} \et. : 2003,
\newblock {\em GRB Coordinates Network} 2459

\bibitem{2007A&A...466...11G}
J.~{Guy} \et. : 2007,
\newblock {\em \aap} 466, 11--21

\bibitem{2005GCN..3914....1H}
J.~{Haislip} \et. : 2005,
\newblock {\em GRB Coordinates Network} 3914

\bibitem{Hamuy-Calan96}
M.~{Hamuy} \et. : 1996,
\newblock {\em \aj} 112, 2391

\bibitem{2007Icar..189..151K}
J.~{Kubica} \et. : 2007,
\newblock {\em Icarus} 189, 151--168

\bibitem{2006ApJ...644L..63K}
S.~R. {Kulkarni} and A.~{Rau}. : 2006,
\newblock {\em \apjl} 644, L63--L66

\bibitem{1986ApJ...304....1P}
B.~{Paczynski}. : 1986,
\newblock {\em \apj} 304, 1--5

\bibitem{1991ApJ...371L..63P}
B.~{Paczynski}. : 1991,
\newblock {\em \apjl} 371, L63--L67

\bibitem{2006PASP..118.1407P}
D.~L. {Pollacco} \et. : 2006,
\newblock {\em \pasp} 118, 1407--1418

\bibitem{2005Natur.438.1132R}
A.~{Rest} \et. : 2005,
\newblock {\em \nat} 438, 1132--1134

\bibitem{2007ApJ...659...98R}
A.~G. {Riess} \et. : 2007,
\newblock {\em \apj} 659, 98--121

\bibitem{2007arXiv0708.2750S}
M.~{Sako} \et. : 2007,
\newblock {\em ArXiv e-prints} 0708.2750

\bibitem{2007arXiv0704.0655S}
B.~{Sesar} \et. : 2007,
\newblock {\em \aj} 134, 2236--2251

\bibitem{2004MNRAS.347...36S}
A.~J. {Storkey} \et. : 2004,
\newblock {\em \mnras} 347, 36--51

\bibitem{2004ApJ...612..877U}
R.~R. {Uglesich} \et. : 2004,
\newblock {\em \apj} 612, 877--893

\bibitem{2007astro.ph..1041W}
W.~M. {Wood-Vasey} \et. : 2007,
\newblock {\em \apj} 666, 694--715

\end{thebibliography}
%\bibliographystyle{plain}
%\input{ms.bbl}

\end{document}